\documentclass[12pt]{iopart}
\usepackage{epsfig}
\newcommand{\J}{\mathsf J}
\newcommand{\bu}{\bi u}
\begin{document}
\jl{1}
\title{Time discretization of functional integrals}
\author{J H Samson\footnote[1]{Electronic address: 
j.h.samson@lboro.ac.uk}}
\address{Department of Physics, Loughborough University, 
Loughborough, Leics LE11 3TU, United Kingdom}
\begin{abstract}
Numerical evaluation of functional integrals usually involves a finite
($L$-slice) discretization of the imaginary-time axis.  In the
auxiliary-field method, the $L$-slice approximant to the density
matrix can be evaluated as a function of inverse temperature at any
finite $L$ as $\hat{\rho}_{L}(\beta)=[\hat{\rho}_{1}(\beta/L)]^{L}$,
if the density matrix $\hat{\rho}_{1}(\beta)$ in the static
approximation is known.  We investigate the convergence of the
partition function $Z_L(\beta)\equiv\Tr\hat{\rho}_{L}(\beta)$, the
internal energy and the density of states $g_{L}(E)$ (the inverse
Laplace transform of $Z_{L}$), as $L\rightarrow\infty$.  For the
simple harmonic oscillator, $g_{L}(E)$ is a normalized truncated
Fourier series for the exact density of states.  When the
auxiliary-field approach is applied to spin systems, approximants to
the density of states and heat capacity can be negative.  Approximants
to the density matrix for a spin-$1/2$ dimer are found in closed form
for all $L$ by appending a self-interaction to the divergent Gaussian
integral and analytically continuing to zero self-interaction. 
Because of this continuation, the coefficient of the singlet
projector in the approximate density matrix can be negative.  For a
spin dimer, $Z_{L}$ is an even function of the coupling constant for
$L<3$: ferromagnetic and antiferromagnetic coupling can be
distinguished only for $L\ge 3$, where a Berry phase appears in the
functional integral.  At any non-zero temperature, the exact partition
function is recovered as $L\rightarrow\infty$.
\end{abstract}

\submitted
\maketitle

\section{Introduction} \label{intro}
Functional integration is a long-established technique in quantum
mechanics \cite{FH}.  More recently, advances in computing power have
allowed direct Monte Carlo evaluation of such integrals for many-body
systems \cite{vdL}.  Such algorithms are often based on an
auxiliary-field functional integral, which is used in areas as diverse
as strongly correlated electron systems \cite{ESW}, spin systems
\cite{LO} and nuclear structure \cite{Negele}.  The statistical
mechanics of a many-body system on a $d$-dimensional lattice is mapped
onto that of a classical field ${\bi u}({\bi r},\tau)$ (the
\emph{auxiliary field}) in a $d+1$-dimensional slab of extension
$0\le\tau\le\beta=1/kT$ in the imaginary time dimension.  The
many-body system reduces to a system of non-interacting particles
moving in a time-dependent auxiliary field.  To evaluate the integral
over all time evolutions, it is necessary to sample the field at a
finite number $L$ of imaginary times --- not necessarily uniformly or
deterministically spaced --- and extrapolate to the continuum limit
$L\rightarrow\infty$.  The $L=1$, or \emph{static}, approximation maps
the system onto classical statistical mechanics in $d$ dimensions. 
The ground state in this approximation for many-fermion systems is a
single Slater determinant, typically corresponding to the Hartree-Fock
solution; for spin models it is the mean-field ground state.  For
$L>2$, closed paths may enclose an area, breaking time-reversal
invariance and thereby contributing a sign or Berry phase factor to
the integral.  This factor has some important consequences.  It
restores quantization: correlation between phases on neighbouring
sites discriminates between the classically equivalent ferromagnets
and unfrustrated antiferromagnets \cite{FrSt}.  The large-$L$ limit
must also restore symmetry if the auxiliary fields do not have the
full local symmetry (such as in the Ising decomposition of the Hubbard
model) \cite{LH}.  On the other hand, the resulting rapid oscillation
of the integrand (the notorious sign problem) seriously restricts
convergence of Monte Carlo simulations at low temperatures.  The
present author has shown how the distributions of the auxiliary fields
tend to the appropriate quantum distribution (the Wigner function)
with increasing $L$, while numerical convergence becomes increasingly
problematical \cite{S32}.  For 
repulsive interactions, an imaginary auxiliary field is required, 
resulting in a sign problem even in the static approximation.

Since numerical studies of the auxiliary-field functional integral are
frequently hampered by the sign problem, it is of value to investigate
toy models in which the finite-$L$ approximants may be evaluated in
closed form.  The present work is a framework for discussion of these
approximants, specifically for simple spin systems.  This differs from
finite-size scaling in real space; while a lattice truncated in real
space is a cluster, and therefore physically realizable, the
time-discretized system may possess unphysical properties vanishing
only in the continuum limit.  Indeed, in a number of examples the
static approximants to the heat capacity and density of states are not
positive-definite \cite{S10}.  The static approximation (and other
finite-$L$ approximations) give a saddle-point approximation, usually a
variational overestimate, of the ground state energy, but are correct
in the high-temperature limit.  The heat capacity shows competition
between the recovery of quantum fluctuations, which give a negative
contribution at low temperatures, and true thermal fluctuations, which 
give a positive contribution  (exponentially small if there is a 
gap).

\begin{figure}
\begin{center}
\epsfig{file=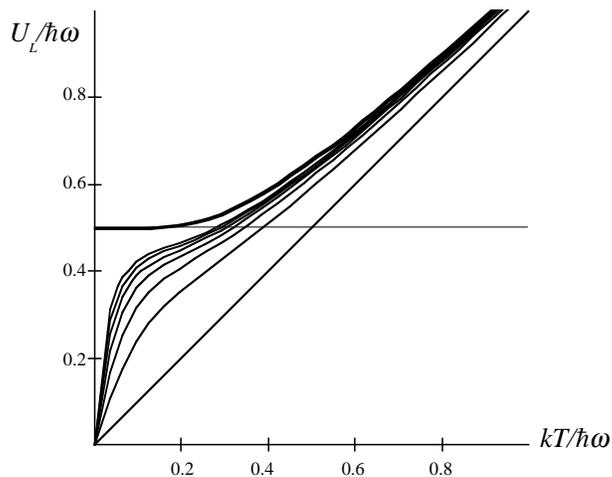,width=8cm} 
\end{center}
\caption{Internal energy for simple harmonic oscillator
showing approximants with $L = 1, 3, 5, 7, 9, 11, 13$
Matsubara frequencies converging to the exact (bold) energy.
\label{fig:SHO}}
\end{figure}

To motivate this work, at this point we recall the
path integral of a simple harmonic oscillator in the frequency domain
(which does not suffer from the above problem).  The partition
function is \cite{FH}
\begin{equation}
    Z(\beta) = \int {\cal D}x \exp\left(-\int_{0}^{\beta}\rmd\tau 
    \left[\frac{m}{2\hbar^{2}}\left(\frac{\rmd x}{\rmd \tau}\right)^{2}
    +\frac{1}{2}m\omega^{2}x^{2}\right]\right).
    \label{eq:ZSHO}
\end{equation}
We impose a frequency cutoff,  restricting the function space to paths
with $L$ Matsubara frequencies ($L$ odd):
\begin{equation}
    x(\tau) = \sum_{n=(1-L)/2}^{(L-1)/2}a_{n} \rme^{2\pi\rmi nkT\tau}.
    \label{eq:Matsubara}
\end{equation}
The resulting $L$th approximant to the partition function is 
\cite{Kleinert,GS}
\begin{equation}
    Z_{L}(\beta) 
    = \frac{1}{\beta\hbar\omega}\prod_{n=1}^{(L-1)/2} \left[ 1 + 
    \left(\frac{\beta\hbar\omega}{2\pi n}\right)^{2}\right]^{-1},
    \label{eq:ZLSHO}
\end{equation}
with poles at $\beta = 2\pi\rmi n/\hbar\omega, (1-L)/2 \le n \le 
(L-1)/2$.  The inverse Laplace transform of $Z_{L}$ gives the $L$th approximant to the 
density of states:
\begin{equation}
    g_{L}(E) = \frac{2^{L-1}\left([(L-1)/2]!\right)^{2}}{(L-1)!\hbar\omega} 
    \sin^{L-1}(\pi E/\hbar\omega)\Theta(E),
    \label{eq:gLSHO}
\end{equation}
where $\Theta$ is the Heaviside step function.  The approximants have
the following limits:
\begin{eqnarray}
    g_{1}(E) & = & \Theta(E)
    \label{eq:g1SHO}  \\
    \lim_{L\rightarrow\infty} g_{L}(E) & \rightarrow & 
    \sum_{n=0}^{\infty}\delta(E-(n+1/2)\hbar\omega)
    \label{eq:ginfSHO}  \\
    g_{L}(E) & \sim
    & E^{L-1}, E\rightarrow 0_{+}.
    \label{eq:g0SHO}
\end{eqnarray}
In particular, \eref{eq:ginfSHO} verifies the emergence of the correct
density of states in the continuum limit.  \Fref{fig:SHO} shows the
convergence of the internal energy $U_{L}$, obtained from $Z_{L}$
\eref{eq:ZLSHO}, to the exact result
$(\hbar\omega/2)\mbox{coth}(\hbar\omega/2kT)$ for any fixed positive
temperature.  The ground state energy vanishes for all $L$, and the
low-temperature heat capacity is $Lk$; the zero-point energy is
recovered with increasing temperature.  In this case the finite-$L$
approximants to the partition function represent physically realizable
systems (ensembles of harmonic oscillators of the same frequency with
a distribution of energy shifts).  This is to be compared with the 
results to be shown in \fref{fig:1spin} and \fref{fig:2spin}, which do 
not exhibit this behaviour.

Of more relevance to the present paper would be a time discretization
of the path integral \eref{eq:ZSHO}.  The resulting approximants to
the partition function have a similar form to \eref{eq:ZLSHO},
although the poles are non-uniformly spaced \cite{Kleinert,VS,KTL}. 
The approximant to the density of states has a less transparent form
than \eref{eq:gLSHO}, being quasiperiodic rather than periodic, but
still converges with $L$ in any finite energy interval.

\Sref{Theory} presents the auxiliary field formalism used in 
this work.  Two case studies of toy spin models in \sref{Examples} show how truncation 
of the functional integral gives a sequence of approximants which, 
although convergent onto the correct value, do not themselves 
represent any physical system.  \Sref{Discussion} discusses possible 
wider applicability of the features of these models.

\section{Theory} \label{Theory}

The general Hamiltonian with two-body interactions is of the form
\begin{equation}
    \hat{H} = -\sum_{\mu=1}^{N} (K_{\mu}\hat{A}_{\mu}+ 
    K_{\mu}^{*}\hat{A}_{\mu}^{\dag}) - 
    \sum_{\mu,\nu=1}^{N}J_{\mu\nu}\hat{A}_{\mu}^{\dag}\hat{A}_{\nu}.
    \label{eq:H}
\end{equation}
Here ${\hat{\bi A}}=\{\hat{A}_{\mu},\mu=1\ldots N\}$ are
single-particle operators generating a closed algebra; they will be
spin operators in the examples studied here, but might, for example, represent
hopping or pairing operators,
$c_{i\uparrow}^{\dag}c_{j\uparrow}$ or
$c_{\mathbf{k}\uparrow}^{\dag}c_{\mathbf{k}\downarrow}^{\dag}$.  To
avoid notational complications, we assume the
operators to be Hermitian.  This can be achieved by changing the basis
to $\hat{A}_{\mu}^{\dag}+\hat{A}_{\mu}$ and
$\rmi(\hat{A}_{\mu}^{\dag}-\hat{A}_{\mu})$.  To obtain the functional 
integral, we separate the density matrix into 
$L$ time slices,
\begin{equation}
    \hat{\rho}(\beta)\equiv e^{-\beta\hat{H}}=
    \left(e^{-\beta\hat{H}/L}\right)^{L}
    \label{eq:rho}
\end{equation}
and apply the Hubbard-Stratonovich
transformation to each time slice:
\begin{equation}
\fl    \exp(-\beta{\hat H}/L) = \frac{\int
        \rmd^{N}\bu \exp(-\beta
        \bu\cdot \J^{-1}\bu/4L)
        \exp(\beta({\bi K}+\bu)\cdot{\hat{\bi A}}/L)}{\sqrt{\det(4\pi 
        \J L/\beta)}} + \Or(L^{-2}).
    \label{eq:HST}
\end{equation}
This is a formal expression, convergent only for a positive interaction matrix 
$\J$.  In general, one needs to reduce the matrix into positive, zero 
and negative blocks and treat each separately, omitting the auxiliary 
fields in the zero block and using an imaginary 
auxiliary field in the negative block \cite{S27}. An alternative is to add a 
multiple of a positive matrix to $\J$.  In the latter case, this 
addition may correspond to a constant or one-body term, which can be 
absorbed into $\bi K$, at the cost of introducing a fictitious 
self-interaction; the functional integration will have to work harder 
to remove this self-interaction.  The coefficient of 
the additional term may be analytically continued or extrapolated to 
zero \cite{ADKLO}; it is this approach we shall use here.
The $L$th approximant to the density matrix is then
\begin{equation}
    \hat{\rho}_{L}(\beta)  =   \prod_{n=1}^{L}\frac{\int\rmd^{N}\bu_{n}
         \exp(-
         \beta\bu_{n}\cdot \J^{-1}\bu_{n}/4L)
        \exp( \beta({\bi K}+\bu_{n})\cdot{\hat{\bi A}}/L)}
	{\sqrt{[\det(4\pi L \J/\beta)]}}.
    \label{eq:rhoL}
\end{equation}
This can be obtained from the density matrix in the static 
approximation
\begin{equation}
    \hat{\rho}_{L}(\beta)  =  [\hat{\rho}_{1}(\beta/L)]^{L},
    \label{eq:rho1} 
\end{equation}
and tends to the exact density matrix as $L\rightarrow\infty$.  We
shall subsequently refer to $L$th approximant of ``function''
(obtained by replacing the exact density matrix with its approximant
\eref{eq:rhoL}) as the $L$-``function''.

The $L$-partition function is
\begin{equation}
    Z_{L}(\beta) = \Tr \hat{\rho}_{L}(\beta).
    \label{eq:ZL}
\end{equation}
Approximants to the internal energy may be computed directly 
from approximants to the partition function,
\begin{equation}
    U_{L}(\beta)  =  -\frac{\partial}{\partial\beta}\ln 
        Z_{L}(\beta),
        \label{eq:U1}
\end{equation}
which is an average of the one-body 
(auxiliary-field) Hamiltonian:
\begin{eqnarray}
	 \fl && U_{L}(\beta)  = - \frac{L}{2}kT + \\
	 \fl &&\frac{\Tr\left\{\int\prod_{n=1}^{L}\left[\rme^{-\beta\left(\bu_{n}\cdot
	      \J^{-1}\bu_{n}/4-({\bi K}+\bu_{n})\cdot{\hat{\bi A}}
	      \right)/L}\rmd^{N}\bu_{n} \right]\left[\bu_{1}\cdot
	      \J^{-1}\bu_{1}/4-({\bi K}+\bu_{1})\cdot{\hat{\bi A}}\right]\right\}}
	 {\Tr\left\{\int\prod_{n=1}^{L}\left[\rme^{-\beta\left(\bu_{n}\cdot
	 \J^{-1}\bu_{n}/4-({\bi K}+\bu_{n})\cdot{\hat{\bi A}}
	 \right)/L}\rmd^{N}\bu_{n} \right]\right\}}. \nonumber
    \label{eq:U1fi}
\end{eqnarray}
This will therefore tend to the expectation of the auxiliary field 
Hamiltonian at low temperatures, typically a mean field energy.

Since it is difficult to extract the partition function from
importance-sampled Monte Carlo calculations, the form
\eref{eq:U1} is impractical.  It is possible to calculate the energy
as a thermal average of the true Hamiltonian
\begin{equation}
    \tilde{U}_{L}(\beta)  =  Z_{L}(\beta)^{-1}\Tr 
        [\hat{\rho}_{L}(\beta)\hat{H}].
        \label{eq:U2}
\end{equation}
This is a variational approximation to the ground 
state energy, which might be expected to be bounded below by the ground state energy.  
The example in \sref{dimer} below shows that this natural assumption is not 
always justified for these approximants.  The forms \eref{eq:U1} and 
\eref{eq:U2} are not equivalent; the latter is usually a better 
approximation.  Heat capacities will be defined as temperature 
derivatives of these energies, although these may be calculated in 
other ways \cite{FS}.

The partition function is the Laplace transform of the density of 
states $g_{L}(E)$.  The $L$-density of states, 
$g_{L}(E)$, is defined implicitly by
\begin{equation}
    Z_{L}(\beta) = \int_{-\infty}^{\infty}e^{-\beta E}g_{L}(E) \rmd E.
    \label{eq:Laplace}
\end{equation}
The spectrum is bounded below but, if necessary, the origin of $E$ 
can be shifted to ensure that $g_{L}(E)=0$ for $E<0$.  Such densities 
of states have been studied in the nuclear shell model \cite{NA}, 
although in that case the partition function is derived by integration 
of the measured energy \eref{eq:U2} in \eref{eq:U1} and the inverse Laplace transform is computed 
within the 
saddle-point approximation (which is appropriate for a large 
density of states).  The propagator may also be determined by 
a similar inverse transform of the density matrix.  The $L$-partition 
function 
$Z_{L}(\beta)$ converges pointwise to the partition function 
$Z(\beta)$ as $L\rightarrow\infty$ at any non-zero temperature.  The 
$L$-density of states converges
to the true density of states
in the distributional sense: for any sufficiently smooth 
function $f$,
\begin{equation}
    \lim_{L\rightarrow\infty}\int_{-\infty}^{\infty}f(E)g_{L}(E)\rmd 
    E = \int_{-\infty}^{\infty}f(E)g(E)\rmd E.
    \label{eq:distconv}
\end{equation}
There are now two possibilities.  If
$\int_{-\infty}^{\infty}f(E)g_{L}(E)\rmd E$ is positive for all 
positive test functions $f(E)$, the heat capacity is non-negative 
at all temperatures and we say that the approximant is physical; there can 
exist an Hermitian Hamiltonian with that thermodynamics.  This is 
evidently the case for the harmonic 
oscillator discussed in \sref{intro}, although this is not 
related to the auxiliary-field functional integral.  If the density of 
states is non-positive, then we say the approximant is unphysical.  The spin 
models in the next section provide examples.

\section{Examples}
\label{Examples}
\subsection{Single spin}
\begin{figure}
\begin{center}
\epsfig{file=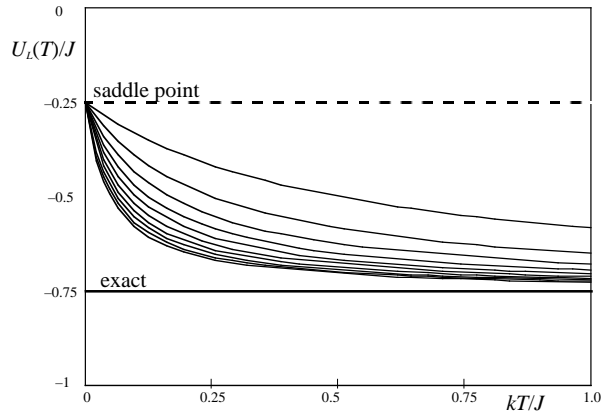,width=8cm} 
\end{center}
\caption{Internal energy for single spin $1/2$ showing approximants
with $L = 1$ (top curve) to 10 (bottom curve) converging to the exact
(bold) energy.
\label{fig:1spin}}
\end{figure}

A single spin $s$ with self-interaction,
\begin{equation}
    \hat{H} = -J \hat{\bi S}\cdot\hat{\bi S},
    \label{eq:H1spin}
\end{equation}
may seem a trivial case, although a similar situation would arise in the
study of a Hubbard model with degenerate bands and strong Hund's rule
coupling.  The static approximation to this has been discussed
earlier \cite{S10}.  Although a scalar auxiliary field does not suffer 
from this problem, it violates rotational invariance 
\cite{Kakehashi}.  Clearly the exact partition function, internal
energy and density of states are 
\begin{eqnarray}
    Z(\beta) & = & (2s+1)\exp(\beta Js(s+1))
    \label{eq:Z1exact}  \\
    U(\beta) & = & -Js(s+1)
    \label{eq:U1exact}  \\
    g(E) & = & (2s+1)\delta(E+Js(s+1)).
    \label{eq:g1exact}
\end{eqnarray}
Applying the 
Hubbard-Stratonovich transformation gives the $L=1$ density matrix as
\begin{equation}
    \hat{\rho}_{1} = (\beta/4\pi J)^{3/2}\int
        \rmd^{3}\bu \exp(-\beta
        u^{2}/4J)
        \exp(\beta\bu\cdot{\hat{\bi S}}).
    \label{eq:HST1spin}
    \end{equation}
This is rotationally invariant, and therefore a multiple of the unit 
matrix.  The partition function in the static approximation follows 
from taking the trace of the exponential and performing the Gaussian 
integrals 
\cite{S10}, giving
\begin{equation}
    \hat{\rho}_{1}(\beta) = \frac{Z_{1}(\beta)}{2s+1} =
    \frac{1}{2s+1}\sum_{m=-s}^{s}(1+2m^{2}\beta J) \rme^{m^{2}\beta
    J}.
    \label{eq:rho1spin}
\end{equation}
The $L$-partition function is then
\begin{equation}
    Z_{L}(\beta) = (2s+1)\left(\frac{1}{2s+1}\sum_{m=-s}^{s}(1+2m^{2}\beta 
    J/L) \rme^{m^{2}\beta
    J/L}\right)^{L};
    \label{eq:ZL1spin}
\end{equation}
this tends to the correct limit \eref{eq:Z1exact} as
$L\rightarrow\infty$.  The $L$-energy \eref{eq:U1} is
\begin{equation}
    U_{L}(\beta) = -J\frac{\sum_{m=-s}^{s}(3m^{2}+2m^{4}\beta 
    J/L) \rme^{m^{2}\beta
    J/L}}{\sum_{m=-s}^{s}(1+2m^{2}\beta 
    J/L) \rme^{m^{2}\beta
    J/L}} = U_{1}(\beta/L).
    \label{eq:UL1spin}
\end{equation}
This is a monotonically \emph{decreasing} function of temperature, 
falling from the saddle-point value of $-Js^{2}$ at $T=0$ to the 
correct value of $-Js(s+1)$ at high temperatures.  \Fref{fig:1spin} 
shows the energy for spin $1/2$.  The thermal average of the Hamiltonian \eref{eq:U2} is trivially 
$\tilde{U}_{L}(\beta)=-Js(s+1)$ at all temperatures.

This negative heat capacity implies a non-physical density of states.  
For spin $1/2$, the approximants to the partition function are
\begin{equation}
    Z_{L}(\beta) = 2(1+\beta J/2L)^{L}\rme^{\beta J/4},
    \label{eq:ZLspinhalf}
\end{equation}
converging to $2\rme^{3\beta J/4}$ as $L\rightarrow\infty$.  Its 
inverse Laplace transform is the $L$-density of states
\begin{equation}
    g_{L}(E) = 2\left(1+\frac{J}{2L}\frac{\rmd}{\rmd E}\right)^{L} 
    \delta(E+J/4),
    \label{eq:gL1spin}
\end{equation}
which involves $L$ derivatives of the 
delta function at the saddle-point energy.  The formal limit
\begin{equation}
    \lim_{L\rightarrow\infty}g_{L}(E) =
    2\exp\left(1+\frac{J}{2}\frac{\rmd}{\rmd E}\right) \delta(E+J/4) =
    2\delta(E+3J/4)
    \label{eq:gLispinlim}
\end{equation}
is correct when applied to a sufficiently good test function (such as
a polynomial).  In this way expectation values are correct to
$\Or(L^{-1})$, even though the only spectral point is at the classical
rather than the quantum ground state energy, and is in error by
$\Or(1)$.  For larger spins, this single
singularity becomes a discrete spectrum of singularities between
$E=-Js^{2}$ and $E=-J/4$ (half-odd-integer spin) or $E=0$ (integer
spin), all higher than the true energy.

There is a suggestive but probably fortuitous resemblance between the partition 
function for spin $1/2$ and the $q$-Laplace 
transform, recently-introduced in the context of non-extensive 
statistical mechanics \cite{LBM}.  One version of this
$q$-Laplace transform defines a $q$-partition function as
\begin{equation}
    Z_{q}(\beta) = \int_{0}^{\infty}g(E)[1+(1-q)\beta E]^{1/(1-q)}\rmd 
    E,
    \label{eq:qLaplace}
\end{equation}
where $q$ corresponds to $1-1/L$.  Negative heat capacities 
are found in this theory \cite{Abe} (although the correspondence 
between $Z_{q}$ and thermodynamic potentials differs from that in 
standard thermodynamics).

\subsection{Spin $1/2$ dimer} \label{dimer}
The highly non-physical behaviour of the approximants to the heat 
capacity above stems from the emergence of the quantum 
fluctuations (a negative energy contribution) with increasing 
temperature.  For a single spin in zero field there are no 
compensating thermal fluctuations.  We therefore investigate the 
spin-$1/2$ dimer,
\begin{equation}
    \hat{H} = -J'(\hat{\bi S}_{1}\cdot\hat{\bi S}_{1} + 
    \hat{\bi S}_{2}\cdot\hat{\bi S}_{2}) 
    - 2J\hat{\bi S}_{1}\cdot\hat{\bi S}_{2}.
    \label{eq:H2spin}
\end{equation}
The self-interaction is added to ensure convergence.  The
integral \eref{eq:HST} only converges for $J'>|J|$, but is analytic in
the matrix elements, allowing continuation to $J'=0$.  Manipulation of the
Gaussian integrals eventually gives the $L$-density
matrix as
\begin{eqnarray}
\fl \hat{\rho}_{L}(\beta) & = & \left[\left(
    \frac{5}{6}-\frac{J'^{2}}{6J^{2}}+
    \frac{\beta (J+J')^{2}}{3LJ}\right)\rme^{\beta J/2L}+
    \left(\frac{1}{6}+\frac{J'^{2}}{6J^{2}}-
    \frac{\beta (J-J')^{2} }{6LJ}\right)\rme^{-\beta J/2L} 
    \right]^{L}e^{\beta J'/2} \hat{P}_{1}
    \nonumber \\
\fl &+& \left[ \left(-\frac{1}{2}+\frac{J'^{2}}{2J^{2}}\right)e^{\beta J/2L}+
    \left(\frac{3}{2}-\frac{J'^{2}}{2J^{2}}
    -\frac{\beta (J-J')^{2}}{2LJ}\right)\rme^{-\beta J/2L}
    \right]^{L}e^{\beta J'/2} \hat{P}_{0}.
    \label{eq:rhoL2spinJJ}
\end{eqnarray}
This is an entire function of both $J$ and $J'$.  Taking 
$J\rightarrow 0$ gives the direct product of two one-particle density 
matrices \eref{eq:ZL1spin}.  More importantly, we can remove the 
interaction by setting $J'=0$ to obtain
\begin{eqnarray}
    \hat{\rho}_{L}(\beta) & = & \left[\left(
    \frac{5}{6}+\frac{\beta J}{3L}\right)\rme^{\beta J/2L}+
    \left(\frac{1}{6}-\frac{\beta J}{6L}\right)\rme^{-\beta J/2L} 
    \right]^{L} \hat{P}_{1}
    \nonumber \\
    &+& \left[ -\frac{1}{2}\rme^{\beta J/2L}+
    \left(\frac{3}{2}-\frac{\beta J}{2L}\right)\rme^{-\beta J/2L}
    \right]^{L} \hat{P}_{0}
    \label{eq:rhoL2spin}
\end{eqnarray}
where $\hat{P}_{1}$ and $\hat{P}_{0}$ are projections onto the triplet 
and singlet subspace respectively.
In the large-$L$ limit we recover the correct density matrix:
\begin{eqnarray}
    \lim_{L\rightarrow\infty}\hat{\rho}_{L}(\beta) & = &  
    \lim_{L\rightarrow\infty}
    \left[\left(1+\frac{\beta J}{2L}\right)^{L} \hat{P}_{1}
    +\left(1-\frac{3\beta J}{2L}\right)^{L} \hat{P}_{0}\right] \\
     & = & \rme^{\beta J/2}\hat{P}_{1} 
     +\rme^{-3\beta J/2} \hat{P}_{0}.
    \label{eq:rhoL2spinlim}
\end{eqnarray}
   
It is not possible to distinguish ferromagnetic from antiferromagnetic
coupling in the thermodynamics for $L<3$, where the paths do not
enclose an area.  Thus the $L$-partition function is an
even function of $J$ for $L=1, 2$.  A high-temperature expansion 
(Maple) verifies this and shows that the second moment of the density of
states is correct for all $L$:
\begin{equation}
\fl Z_{L}(\beta) = 4 + \frac {3(\beta J)^{2}}{2} - \frac {(L - 1)(L -
 2)(\beta J)^{3}}{2L^{2}} + \frac { (21L^{3}- 72L^{2} +
 116L - 60) (\beta J)^{4}}{96L^{3}} + \cdots.
 \label{eq:Zexpand}
\end{equation}

\begin{figure}
\begin{center}
\epsfig{file=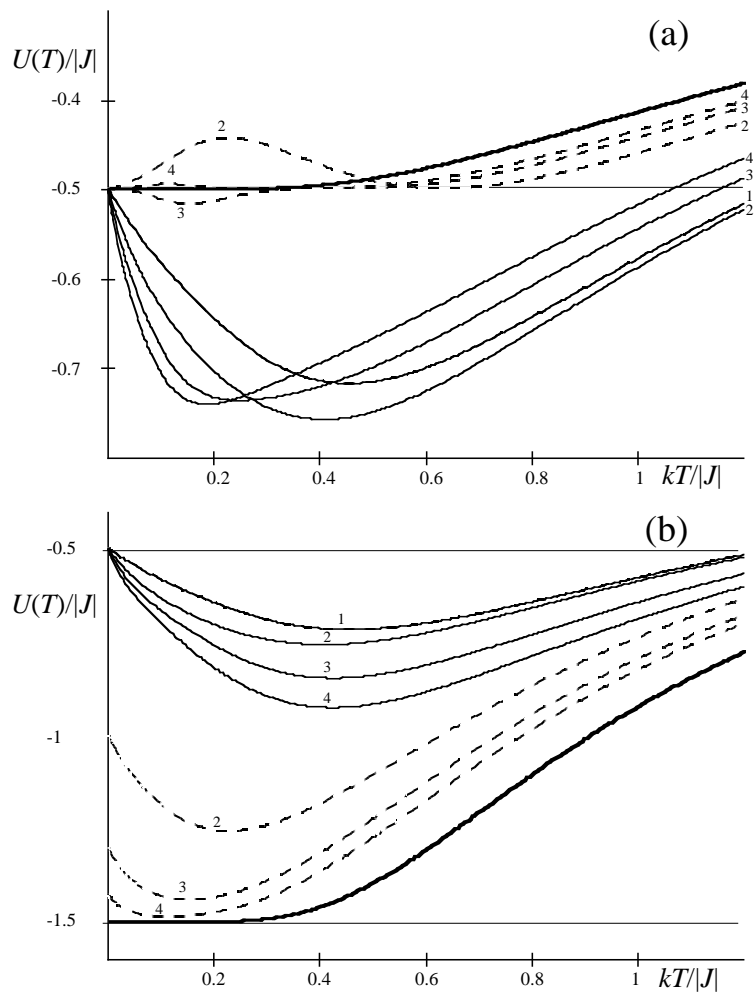,width=10cm} 
\end{center}
\caption{$L$th approximant to the energy for spin-$1/2$ dimer with (a) 
ferromagnetic and (b) antiferromagnetic coupling.  The bold line is 
the exact energy, the full lines are $U_{L}$, the derivative of 
$\ln Z_{L}$ \eref{eq:U1}, and the dashed lines are $\tilde{U}_{L}$, the 
thermal average of the Hamiltonian \eref{eq:U2}.  Curves shown for $L=1\ldots 4$, 
annotated by $L$.  The energies $U_{1}$ and $\tilde{U}_{1}$ are coincident.
\label{fig:2spin}}
\end{figure}

\Fref{fig:2spin} shows approximants to the internal energy for
ferromagnetic and antiferromagnetic coupling.  $U_{L}$, as calculated
from the partition function \eref{eq:U1}, is always equal to its mean
field value $-|J|/2$ at $T=0$.  This is the correct energy only for
the ferromagnet.  $U_{1}$ and $U_{2}$, as already discussed, cannot
distinguish the ferromagnet and antiferromagnet.  The $L$-heat
capacity is negative at low temperatures and positive at higher
temperatures.  $\tilde{U}_{L}$, as calculated from the thermal average
of the Hamiltonian \eref{eq:U2} is, as expected, a better
approximation than $U_{L}$ for $L>1$, giving improved estimates of the
antiferromagnetic ground state energy, although it still shows a
small region of negative $L$-heat capacity.  

One at first surprising feature is that in the ferromagnet
$\tilde{U}_{L}$ falls \emph{below} its variational bound $-J/2$ for
odd $L$.  This is due to the unphysical form of the $L$-density matrix
itself, and not just to its temperature dependence.  As a result of
the analytic continuation to $J'=0$, the coefficient of the singlet
projector in \eref{eq:rhoL2spin} is negative at low temperature for
odd $L$, representing a negative weight for the singlet state.  The
coefficients of both projectors are always positive when $J'>|J|$, the
parameter region for which the integral \eref{eq:HST} converges. 
Direct application of the Hubbard-Stratonovich transformation to the
Hamiltonian for $J'=0$ would require an imaginary field coupled to
$\hat{\bi S}_{1}-\hat{\bi S}_{2}$, leading to a similar non-classical
weight.

The $L$-density of states is again non-physical and is symmetric for 
$L<3$ for the reasons discussed above; there are $L+1$ singularities in $-J/2\le E \le J/2$, involving $L$
derivatives of the delta function, for example
\begin{equation}
    g_{1}(E)  =  
           2\left(\delta(E+J/2)+\delta(E-J/2)\right)
	   +J\left(\delta'(E+J/2)-\delta'(E-J/2)\right).
    \label{eq:g1dimer}
\end{equation}
For large $L$, we obtain the
correct result (the limit to be understood in the
distributional sense)
\begin{equation}
     \lim_{L\rightarrow\infty}g_{L}(E) = 3\delta(E+J/2) +
	 \delta(E-3J/2).
    \label{eq:ginfdimer}
\end{equation}

\section{Discussion}\label{Discussion}
In all the above, the approximants to thermodynamic functions have 
error $\Or(L^{-1})$ at any fixed non-zero temperature.  In practice, 
more careful Trotter decompositions and truncations of the density matrix 
may accelerate the convergence in Monte Carlo simulations 
\cite{FS,Suzuki,Fye}. 
The models discussed, being analytically soluble for all 
discretizations, are not representative of real applications but may 
provide a useful test of methods.  The main outcome of this work is a 
pointer to possible difficulties in the use of finite discretizations: 
underestimated (or negative) heat capacities and non-physical spectral 
functions.  The effects might be largest in strongly correlated 
systems, or systems with an excitation gap, where the lowest 
auxiliary field state is a poor approximation to the true ground 
state.  In that case the density of states must suffer substantial 
distortion to provide the correct thermodynamics; the true ground 
state energy lies outside the approximate spectrum.  More accurate 
energies are obtained from the thermal average of the Hamiltonian than 
from the derivative of the partition function.

This work has been in some sense complementary to the sign problem,
which can arise for $L\ge 3$ or for repulsive interactions \cite{S27}:
the weight in the functional integral \eref{eq:HST} need not be
positive, although physical quantities are correctly obtained.  This
will arise if the correlations to be calculated are incompatible with
a positive distribution for the auxiliary fields \cite{S32}.  In this case low-$L$
approximations lead to unphysical results, characterized by
non-positive distributions in the energy domain.  The
examples discussed are those with the most acute sign problem: the
auxiliary field couples to operators (such as spin components) whose
equal-time commutators do not vanish.

\section*{References}


\end{document}